\documentclass{article}

\usepackage[square,numbers]{natbib}
\usepackage{times}
\usepackage{amsfonts}
\usepackage{xcolor}
\usepackage[linesnumbered, ruled, vlined, algo2e]{algorithm2e}
\usepackage{layout}

\usepackage[dvipsnames]{xcolor}
\definecolor{darkgreen}{rgb}{0,0.5,0}
 
\makeatletter
\@ifundefined{lemma}{}{ }
\@ifundefined{definition}{}{ }
\@ifundefined{assumption}{}{ }
\@ifundefined{corollary}{}{ }
\@ifundefined{remark}{}{ }
\makeatother

\newtheorem{lemma}{Lemma}
\newtheorem{definition}{Definition}   
\newtheorem{corollary}{Corollary}
\newtheorem{remark}{Remark}
\newtheorem{theorem}{Theorem}
\usepackage{graphicx}
\usepackage{tcolorbox}
\AtBeginDocument{%
  \providecommand\BibTeX{{%
    \normalfont B\kern-0.5em{\scshape i\kern-0.25em b}\kern-0.8em\TeX}}}

\usepackage{wrapfig}
\usepackage{booktabs} 

\usepackage{url}
\usepackage{algorithm}
\usepackage[noend]{algpseudocode}
\usepackage{amssymb,amsmath}
\usepackage{subcaption}
\usepackage{comment}

\usepackage{thmtools, thm-restate}
\usepackage{multirow}

\usepackage[graphicx]{realboxes}

\usepackage{xcolor} 
\definecolor{ocre}{RGB}{243,102,25} 

\usepackage{avant} 
\usepackage{mathptmx} 

\usepackage{microtype} 
\usepackage[utf8x]{inputenc} 

\usepackage{tikz}
\usetikzlibrary{shapes}
\RequirePackage{ifthen}
\RequirePackage{amsmath}
\RequirePackage{amssymb}
\RequirePackage{xspace}
\RequirePackage{graphics}
\usepackage{color}
\usepackage{keyval}
\usepackage{xspace}
\usepackage{mathrsfs}
\usepackage{enumitem}

\DeclareMathAlphabet{\mathcal}{OMS}{cmsy}{m}{n}

\newcommand{\dom}{\relax\ifmmode {\mathit{dom}} \else ${\sf dom}$\fi}

\usepackage{setspace}
\usepackage{listings}
\newcounter{assumptions}
\newtheorem{assumption}[assumptions]{Assumption} 
\newcommand{\pf}{\par\noindent{\bf Proof:}~}
\newcommand{\qed}{\hfill{\rule{2mm}{2mm}}\medskip}
\newenvironment{proof}{\pf}{\qed}

\usepackage[normalem]{ulem}

\usepackage{placeins}
\makeatletter
\newcommand\nocaption{%
    \renewcommand\p@subfigure{}
    \renewcommand\thesubfigure{\thefigure\alph{subfigure}}
}
\makeatother

\begin{document}

\title {Safe Control using Learned Safety Filters and Adaptive Conformal Inference}

\author{
	Sacha Huriot, Ihab Tabbara, and Hussein Sibai \\
	{\small Computer Science \& Engineering}\\
	{\small Washington University in St. Louis} \\
	{\small \texttt{\{h.sacha,i.k.tabbara,sibai\}@wustl.edu}}
}

\date{}
\maketitle

\begin{abstract}

Safety filters have been shown to be effective tools to ensure the safety of control systems with unsafe nominal policies. To address scalability challenges in traditional synthesis methods, learning-based approaches have been proposed for designing safety filters for systems with high-dimensional state and control spaces. However, the inevitable errors in the decisions of these models raise concerns about their reliability and the safety guarantees they offer. This paper presents Adaptive Conformal Filtering (ACoFi), a method that combines learned Hamilton-Jacobi reachability-based safety filters with adaptive conformal inference. Under ACoFi, the filter dynamically adjusts its switching criteria based on the observed errors in its predictions of the safety of actions. The range of possible safety values of the nominal policy's output is used to quantify uncertainty in safety assessment. The filter switches from the nominal policy to the learned safe one when that range suggests it might be unsafe. 
We show that ACoFi guarantees that the rate of incorrectly quantifying uncertainty in the predicted safety of the 
nominal policy 
is asymptotically upper bounded by a user-defined parameter.
This gives a soft safety guarantee 
rather than a hard safety guarantee. 
We evaluate ACoFi in a Dubins car simulation and a Safety Gymnasium environment, empirically demonstrating that it significantly outperforms the baseline method that uses a fixed switching threshold by achieving higher learned safety values and fewer safety violations, especially in out-of-distribution scenarios.

\end{abstract}

\textbf{Keywords:}
Conformal prediction, safety filters, safe control\section{Introduction}
Assuring safety is essential for deploying safety-critical control  systems, such as self-driving cars \cite{end_to_end_AD_survey_2024} and surgical robots \cite{autonomy_surgical_robots}. Safety filters are prominent tools for ensuring their safety. Control barrier functions (CBFs) \cite{ames2019control} and Hamilton-Jacobi (HJ) reachability value functions \cite{bansal2017hamilton} have been used to design safety filters that guarantee safe operation of control systems by adjusting their unsafe nominal actions to safe ones. However, traditional methods for synthesizing CBFs, such as sum-of-squares programming \cite{zhang2023efficient,clark2021verification}, and for computing HJ reachability value functions, such as dynamic programming \cite{mitchell2005HJ}, suffer from the curse-of-dimensionality. This motivated data-driven approaches for learning safety filters \cite{HJ_RL,CCBF,lipschitz_HJ,how_to_train_2024,CBF_and_Input_to_State_Safety_for_AD_TCST_2023}.

In our work, without loss of generality, we focus on designing reliable safety filters relying on a learned HJ reachability value function $V_\theta$. An instance of such filters evaluates the safety of the nominal control action at every state reached, and if it considers it unsafe, it switches to the learned safe policy that optimizes $V_\theta$. Importantly, this backup is not assumed to be a perfect safe policy for the true system, but rather the safest policy induced by the current learned approximation of the safety  value function. Our goal is therefore not to construct a perfect safety filter, but to determine when to switch from task execution to using this safest policy available. Existing methods that rely on such safety filters use fixed thresholds for the value functions evaluating the safety of proposed actions to switch between the nominal and learned safe policies. However, when a HJ value function is learned from data, it is not guaranteed to be correct, and it is expected to be more erroneous in regions of the state space that are poorly represented during training, making fixed thresholds unreliable
\cite{chen2018signal,SafetyFrameworkforUncertainty,lin2024filterdeployallrobust,designinglatentsafetyfilters}.

In order to quantify the uncertainties of black-box predictors, conformal prediction \cite{conformal_prediction_original_gammerman_1998, conformal_decision_making_vovk_2018} has emerged as a statistical framework for generating confidence regions called {\em conformal sets}. Given a desired miscoverage rate $\alpha$, calibration data, and a predictor input $x$, the corresponding real output $y$ will belong to  the conformal set generated by conformal prediction with at least $1-\alpha$ probability. This method relies on the exchangeability assumption, i.e., that the joint distribution of the calibration data and the new test point 
is invariant under any permutation. However, the states and actions in trajectories are not exchangeable. Adaptive Conformal Inference (ACI) extends the application of conformal prediction to time-dependent data \cite{adaptive_conformal_inference_under_distribution_shift_gibbs_neurips_2021}. In ACI settings, time series data, such as trajectories of dynamical systems, are considered. At each time step, the black-box predictor predicts the data point in the next time step and then the true data point is observed at that time step, 
i.e., ground-truth is observed in a delayed manner. ACI results in time-dependent conformal sets which guarantee that the average rate of miscoverage over time is bounded by a user-defined parameter $\alpha$.

To address the failure of existing data-driven safety filters in accounting for their prediction errors, we propose Adaptive Conformal Filtering (ACoFi), a method that dynamically adjusts the criteria according to which these filters switch from the nominal policies to the learned safe ones. By monitoring the difference between the learned safety value at the current state and the updated one  after receiving the observation at the next state, ACoFi adapts the threshold for switching from the nominal policy to the learned safe one corresponding to the HJ value function, providing probabilistic guarantees on the average rate of actions taken over time that are deemed unsafe by the learned safety value function, while minimizing unnecessary switching. 
We evaluate ACoFi in two vision-based navigation tasks. We show that ACoFi outperforms fixed threshold-based switching baselines by achieving higher safety values and performing fewer unsafe actions without excessive switching to the learned safe policy. 

Our contributions are: (1) we introduce ACoFi, a method that uses ACI to account for the prediction errors of learned safety filters and provides formal guarantees, and (2) we empirically demonstrate ACoFi's effectiveness  in high-dimensional control settings. 
\subsection{Related work}
\label{sec:relatedwork}
Safe control under uncertainty has been widely explored, particularly for high dimensional systems prone to operating in out-of-distribution (OOD) conditions \cite{seo2025UNISafe,singletary2022safe, sadigh2016safe, wang2025safe, dacs2025robust,2025statistically, hu2025uncertain, michaux2025can}. In~\cite{CDP_ICRA_2025}, we used the theory of conformal decision policies (CDPs)~\cite{lekeufack2024conformal} to account for the uncertainty in the trajectory predictions of other agents in multi-agent environments while using CBFs to maintain collision avoidance. CDPs offer    
deterministic guarantees, in contrast with the probabilistic guarantees of ACI, on the average-over-time of the number of violations of the uncertainty bounds. In that setting, the ground-truth trajectories are observed after one time step, in contrast with the setting of this paper where the state is only partially observed and the ground-truth is not revealed.   
Another close work to ours is UNISafe \cite{seo2025UNISafe}, which extends traditional latent safety filters by accounting for epistemic uncertainty to avoid regions with OOD dynamics, and thus unseen hazards, and provides safety guarantees using conformal prediction. Before deployment, it calibrates an uncertainty threshold via conformal prediction, defining an OOD region in the latent space that is added to the failure set for which the HJ value function is learned. While this approach enables avoidance of previously unknown failures, it can be overly conservative, preventing entering unseen regions even when that is not safety-violating.  
In contrast, our method applies adaptive conformal inference at deployment time, not during training. A Hamilton-Jacobi reachability value function is first learned without modeling uncertainty, then the safety filter is designed by dynamically adjusting the threshold value for switching based on observed prediction errors. 

Moreover, ~\cite{somil2025reachability} train neural control barrier-value functions (CBVFs) and use conformal prediction to expand their level sets. Then, they solve quadratic programs online to 
compute the 
closest safe control to the nominal one. On the other hand,  \cite{somil2024reachability} use conformal prediction to verify super-level sets of learned BRTs. Other existing works that use HJ value functions as safety filters typically switch to the safe controller whenever the predicted HJ value function evaluated at the next time-step is greater than some user-defined threshold,  without formal guidance on how to choose the threshold~ \cite{generalizingsafetycollisionavoidancelatentspace,designinglatentsafetyfilters}.
These methods do not provide formal guarantees on the safety when following the resulting policy that arises from switching between the nominal and the learned HJ-based safe controllers.


\section{Preliminaries}
\label{sec:Preliminaries}

Consider a control system, or agent, operating in an environment described by unobserved states $z\in\mathcal{Z}$, relying on high-dimensional observations $Obs(z)\in\mathcal{X}$ to pick control actions in a control space, a compact set $\mathcal{U} \subset \mathbb{R}^m$, in order to accomplish a task while avoiding a set of unsafe states $\mathcal{Z}_\mathit{unsafe} :=\{z\in\mathcal{Z}\mid l_\mathcal{Z}(z)<0\}$, for some $l_\mathcal{Z}:\mathcal{Z}\to\mathbb{R}$.
An encoder $E_\phi$ can then be used to generate latent states in a low-dimensional space $\mathcal{Y}$. After each observation $x_{t}=Obs(z_{t})$ of the environment, the encoder combines it with the previous latent state $y_{t-1}$ and returns the current one $y_{t} \sim E_\phi(y_{t}\ |\ y_{t-1}, x_{t})$.
Such an encoder is usually trained as a component of a world model~\cite{nav_world_models}.

\subsection{Hamilton-Jacobi value function}
Given a set $\mathcal{F}$ of failure states, a Hamilton-Jacobi reachability value function $V$ and its associated safe policy $\pi^\mathit{safe}$ define the {\em Backward Reachable Tube} (BRT) of $\mathcal{F}$ for the control system. The BRT is the set of states starting from which the system inevitably eventually enter $\mathcal{F}$ using any policy. Moreover, when starting from a state in the complement of the BRT and following $\pi^\mathit{safe}$, the system never reaches $\mathcal{F}$~\cite{bansal2017hamilton}. 
\cite{generalizingsafetycollisionavoidancelatentspace} and \cite{designinglatentsafetyfilters} train a classifier $l:\mathcal{Y}\to\mathbb{R}$ over the latent space that defines the failure set $\mathcal{F}:=\{y\in\mathcal{Y}\mid l(y)<0\}$. Then, they  
conduct approximate HJ reachability analysis in the latent space to train both a HJ reachability value function $V_\theta:\mathcal{Y}\to\mathbb{R}$ and a corresponding safety-preserving policy $\pi^\mathit{safe}_\theta:\mathcal{Y}\to\mathcal{U}$.
For $y \in \mathcal{Y}$, the HJ value function is defined as $V_\theta(y) := \max_{u \in \mathcal{U}} Q_\theta(y, u)$, and the policy as $\pi^\mathit{safe}_\theta(y):=\arg\max_{u \in \mathcal{U}}Q_\theta(y, u)$,
where $Q_\theta$ is the associated Q-function. This Q-function $Q_\theta$ is  learned by employing 
reinforcement learning methods such as DDPG \cite{ddpg} and DDQN \cite{ddqn} to minimize the following loss function: 
\begin{align}
    L(\theta):= \mathbb{E}_{(y_t,u_t,y_{t+1})\sim D}\left[(Q_\theta(y_t, u_t) - R(y_t,u_t,y_{t+1}))^2\right],
\label{eq:safety_q_loss}
\end{align}
where $R$ is the {\em target function} and is defined as follows: 
\begin{align}
    R(y_t,u_t,y_{t+1}) := (1-\gamma)l(y_t) + \gamma \min\left\{l(y_t),\ \max_{u \in \mathcal{U}} Q_\theta(y_{t+1}, u)\right\},\label{eq:safety_q_target}
\end{align}
where $\gamma\in(0,1)$ is a discounting parameter. The policy $\pi^\mathit{safe}_\theta$ can either be computed at runtime when the action space is finite and small or can be a learned along with the Q-function using actor-critic methods, otherwise. One can then plug $\pi^\mathit{safe}_\theta$ in the second argument of $Q_\theta$ to compute $V_\theta$.

\subsection{Adaptive conformal inference}

Consider data points in the form of  $(X,Y)\in\mathbf{X}\times\mathbf{Y}$ sampled from an unknown distribution for some sets $\mathbf{X}$ and $\mathbf{Y}$. Given a predictor $\mu:\mathbf{X}\to\mathbf{Y}$,
the conformal prediction framework uses a {\em calibration dataset} $\{(X_n,Y_n)\}_{n\in[N]}$, a score function $s:\mathbf{Y}^2\to\mathbb{R}$, and a {\em miscoverage rate} $\alpha$, to compute the $(1-\alpha)$-quantile $q$ of the set of conformal scores $\{s(\mu(X_n),Y_n)\}_{n\in[N]}$. If the joint distribution from which the calibration data set and fresh data points are sampled is invariant under permutations, it is called exchangeable. 
In that case, for any freshly sampled data point $(X',Y')$, $Y'$ is guaranteed to belong to the conformal set $I_{N+1}:=\{Y\in\mathbf{Y}\mid s(\mu(X'),Y)\leq q\}$ with probability at least $1-\alpha$ over the joint distribution of the calibration set and the fresh data point~\cite{ConformalPredictionGentleIntro}.   

\par
Adaptive Conformal Inference (ACI) \cite{adaptive_conformal_inference_under_distribution_shift_gibbs_neurips_2021} extends this method to repeated predictions in a time-series $\{(X_{t},Y_{t})\}_{t\in\mathbb{N}_{\geq1}}$, even under distribution shift. In ACI, the true output is observed in a delayed fashion, e.g., at the next step. For step $t\geq1$, the series's history $\{(X_{t'},Y_{t'})\}_{t'<t}$ is considered as the calibration dataset, and an {\em effective miscoverage rate} $\alpha_t$ is used to define the quantile $q_{t}$ of the set of conformal scores. This rate adapts to the observed prediction errors using the {\em update rule}   $\alpha_{t+1}:=\alpha_t+\lambda(\alpha-\mathit{err}_t),$
with a fixed user-defined {\em learning rate} $\lambda$ and {\em target miscoverage rate} $\alpha$. The error term is defined as $\mathit{err}_t:=1[Y_{t}\not\in I_{t}]=1[s(\mu(X_{t}),Y_{t})> q_{t}]$. The target miscoverage rate serves as the limit of the average error rate as stated by the following theorem. 
\begin{theorem}Long-term error rate bound \cite{adaptive_conformal_inference_under_distribution_shift_gibbs_neurips_2021}: \label{thm:conformal_bound}
    Fix a user-defined miscoverage rate $\alpha\in [0,1]$ and a learning rate $\lambda \in \mathbb{R}^{>0}$, and consider the update rule for $\alpha_t$. Then, with probability 1, $\frac{1}{T}\sum_{t=1}^T\mathit{err}_t=\alpha+o(1)$, as $T\to\infty$. More precisely, the following holds:
    $$\forall T\in\mathbb{N},\
\left\vert\frac{1}{T}\sum_{t=1}^T\mathit{err}_t -\alpha\right\vert\leq\frac{\max\{\alpha_1, 1 - \alpha_1\}+\lambda}{T\lambda}=O\left(\frac1T\right),$$
where $\alpha_1$ is the user-initialized value of $\alpha_t$. 
\end{theorem}

\section{Methodology}
\label{sec:algorithm_section}
In this section, we describe  ACoFi  and discuss its guarantees.

\subsection{Safe control while accounting for prediction errors}
The safety constraint the agent aims to maintain for the system is $V_\theta(y_t)>0$. Without uncertainty, the previously described $Q_\theta$ and policy $\pi^\mathit{safe}_\theta$ can be used for runtime safety filtering by considering the value 
of $Q_\theta(y,\pi^\mathit{task}(y))$. This takes the form of a switching strategy, using $\pi^\mathit{safe}_\theta$ at $y$ 
when $Q_\theta(y,\pi^\mathit{task}(y))$ 
is below a fixed user-defined threshold $\epsilon>0$ as follows:
\begin{align*}
    \pi^\mathit{fixed}(y)=&1[Q_\theta(y,\pi^\mathit{task}(y))\geq\epsilon]\cdot\pi^\mathit{task}(y)+1[Q_\theta(y,\pi^\mathit{task}(y))<\epsilon]\cdot\pi^\mathit{safe}_\theta(y).
\end{align*}

However, $Q_\theta(y_{t},u_t)$ is not necessarily equal to the target function value $R(y_t,u,y_{t+1})$ because of generalization errors in in-distribution and out-of-distribution states. 
Our method quantifies how such 
errors affect safety and accounts for them in the switching strategy. 
At step $t+1$, when the new 
latent state $y_{t+1}$ is obtained from the new observation at time $t+1$, the target function value at time $t$   
$R_t=R(y_t,u_t,y_{t+1})$ can be computed. Moreover, 
\begin{gather*}
    R_{t}=(1-\gamma)l(y_{t})+\gamma\min\left\{l(y_{t}),V_\theta(y_{t+1})\right\}\leq(1-\gamma)l(y_{t})+\gamma V_\theta(y_{t+1}).
\end{gather*}
Hence, if 
$u_t$ is chosen so that   
$R_{t}\geq\gamma\epsilon+(1-\gamma)l(y_{t})$,  for some $\epsilon > 0$, then  
$V_\theta(y_{t+1})\geq\frac1\gamma\left(R_{t}-(1-\gamma)l(y_{t})\right)$ $\geq\epsilon$, satisfying the safety constraint.

\subsection{Adaptive conformal filtering}
Since we are only concerned with the uncertainty which negatively affects safety, we use the conformal score $S_{t}=s(Q_\theta(y_{t},u_{t}),R_t)=\max\{Q_\theta(y_{t},u_{t})-R_t,\ 0\}$. Our safety filter, Algorithm \ref{alg:1}, tracks the $(1-\alpha_{t})$-quantile of the score history $\{S_{t'}\}_{t'\leq t}$. This defines the following interval: 
\begin{align*}
    I_{t}&=\{\mathrm{r}\in\mathbb{R}\ |\ s(Q_\theta(y_{t},u_{t}),\mathrm{r})\leq q_t\}=\{\mathrm{r}\in\mathbb{R}\ |\ Q_\theta(y_{t},u_{t})-\mathrm{r}\leq q_t\}=\left[Q_\theta(y_{t},u_{t})-q_t,+\infty\right).
\end{align*}
Then, at the next step, the error term $\mathit{err}_{t}=1[R_t\not\in I_{t}]$ is used to update the effective miscoverage level to $\alpha_{t+1}$. We want to 
pick a control $u_t$ such that the safety constraint $V_\theta(y_{t+1})\geq\epsilon$ is satisfied. Accordingly, our method tests  whether $u_t$ satisfies the Q-value constraint. 

\IncMargin{1em}
\begin{algorithm}
\LinesNumbered
\DontPrintSemicolon
\caption{Adaptive Conformal Filtering (ACoFi) Algorithm}\label{alg:1}
\SetKwFunction{Insert}{Insert}\SetKwFunction{Quantile}{Quantile}\SetKwFunction{StepAndEncode}{StepAndEncode}\SetKwFunction{Terminating}{Terminating}
\SetKwInOut{Input}{input}

\Input{Starting latent state $y_1$, control $u_1$, and miscoverage rate $\alpha_{1}$}
\While{$\neg\ $\Terminating{$y_t$}}{
$y_{t+1}\gets$\StepAndEncode{$y_{t},u_{t}$}\label{ln:encode}\;
$R_{t}\gets (1-\gamma)l_{t} + \gamma \min\left\{l_{t},\ V_\theta(y_{t+1})\right\}$\;
$err_{t}\gets 1[s(Q_\theta(y_{t},u_{t}),R_{t})> q_{t}]$\label{ln:err}\;
$\alpha_{t+1}\gets\alpha_{t}+\lambda(\alpha-err_t)$\label{ln:aci}\;
\Insert{$s(Q_\theta(y_{t},u_{t}),R_{t}),\ \mathcal{S}$}\label{ln:score}\;
$q_{t+1}\gets$\Quantile{$\mathcal{S},1-\alpha_{t+1}$}\label{ln:quant}\;
$l_{t+1}\gets l(y_{t+1})$\;
\eIf{$Q_\theta(y_{t+1},\pi^\mathit{task}(y_{t+1}))\geq q_{t+1}+\gamma\epsilon+(1-\gamma)l_{t+1}$}{
$u_{t+1}\gets\pi^\mathit{task}(y_{t+1})$\label{ln:if}\;
}{

$u_{t+1}\gets\pi^\mathit{safe}_\theta(y_{t+1})$\label{ln:safe_val}\;
}
}
\end{algorithm}
The initialization of the algorithm consists of assigning the values $l(y_1)$, $\emptyset$, and 0, to the variables $l_1$, $\mathcal{S}$, and $q_1$, respectively. The algorithm runs until the agent accomplishes the task or the run ends, which is encoded by $\texttt{Terminating}$. Although this is not required, the first control input $u_1$ should ideally be safe to use in state $y_1$, since there is no prior history to evaluate the accuracy of $Q_\theta$ at the start.
The function $\texttt{StepAndEncode}(y_{t},u_{t})$ allows the system to use control $u_t$, observe $x_{t+1}=Obs(z_{t+1})$, and encode $y_{t+1}\sim E_\phi(y_t,x_{t+1})$. $\texttt{Insert}$ updates $\mathcal{S}$ while maintaining it in sorted order for easy quantile computation. The function $\texttt{Quantile}(\mathcal{S},p)$ returns $0$ for $p<0$, $+\infty$ for $p>\frac{|\mathcal{S}|}{|\mathcal{S}|+1}$, and the $\lceil p\cdot(|\mathcal{S}|+1)\rceil$-th element of $\mathcal{S}$, otherwise. This can be interpreted as follows:
\begin{itemize}
    \vspace{-0.2cm}
    \item If $\alpha_{t+1}>1$, then the target values $\{R_{t'}\}_{t'\leq t}$  have satisfied the inequalities  $Q_\theta(y_{t'},u_{t'})- q_{t'}\leq R_{t'}$ enough times to be confident that the safety estimate is currently accurate, i.e., that $R_{t+1}$ will satisfy the inequality  
    $Q_\theta(y_{t+1}, \pi^\mathit{task}(y_{t+1})
    )\leq R_{t+1}$ with high probability. Hence, the quantile $q_{t+1}$ gets assigned the value $0$. That encourages less conservative control, improving task performance. 
    \vspace{-0.2cm}
    \item If $\alpha_{t+1}<1/(|\mathcal{S}|+1)$, 
    then the target values $\{R_{t'}\}_{t'\leq t}$ have violated the conformal bounds $Q_\theta(y_{t'},u_{t'})- q_{t'}\leq R_{t'}$ enough times to be confident that $Q_\theta(y_{t+1}, \pi^\mathit{task}(y_{t+1}))$ is not a good estimate of the safety of 
    following $\pi^\mathit{task}$. 
    Thus, the algorithm prioritizes safety by
    setting the quantile $q_{t+1}$ to  $+\infty$, ensuring that $err_{t+1}=0$ and ensuring a switch to $\pi^\mathit{safe}$ at the current step.
    \vspace{-0.2cm}
\end{itemize}
Finally, from the test statement at the end, we identify the expression $q_{t+1}+\gamma\epsilon+(1-\gamma)l_{t+1}$ as the {\em adaptive threshold} that the Q-value of the task policy must pass to be used.

\subsection{Conformal safety guarantee}
Our adaptive policy is guaranteed to have its average error rate upper bounded by  $\alpha+o(1)$. This means that, in the long term, $\alpha+o(1)$ is an upper bound on the proportion of steps which followed the nominal policy and resulted in a Q-value under the threshold $q_{t+1}+\gamma\epsilon+(1-\gamma)l_{t+1}$. We formalize this in our main theorem, whose proof is in Appendix~\ref{sec:theorem_proof}, and its corollary.

\begin{theorem}
\label{thm:algorithm_guarantees}
    When using Algorithm \ref{alg:1}, with probability one, the computed bound on the next safety value will hold for a proportion of $1-\alpha+o(1)$ of the history. More precisely, 
    $$\frac{1}{T}\sum_{t=1}^T1\left[V_\theta(y_{t+1})\geq B_{t}\right] \geq1-\alpha+O\left(\frac{1}{T}\right),$$
    where the lower bound $B_{t}:=\gamma^{-1} (Q_\theta(y_{t},u_{t})-q_{t}-(1-\gamma)l_{t})$ can be computed at time step $t$.
\end{theorem}

\begin{corollary}
    The proportion of violations of the constraint $V_\theta(y_{t+1})\geq\epsilon$ by task actions over the whole history is at most $\alpha+o(1)$. Indeed, line \ref{ln:if} of the algorithm forbids task control $\pi^\mathit{task}(y_{t+1})$ that would result in $B_{t+1}\leq \epsilon$. Hence, in the long term, a proportion of at least $1-\alpha$ of the steps using 
    $\pi^\mathit{task}$ 
    will be guaranteed to satisfy $V_\theta(y_{t+1})\geq B_{t+1}\geq\epsilon>0$.
\end{corollary}

ACoFi does not certify that $\pi^\mathit{safe}$
is safe, but it decides when it is preferable to stop following $\pi^\mathit{task}$ and hand control to the former as it is expected to be safer than the latter. In particular, early violations are not ignored by the method. They are the feedback that drives ACoFi to become more conservative than a fixed-threshold switching rule under the same uncertainty. Moreover, observed errors in the value of $V_\theta$ at deployment can be assumed to correlate with changes in ground truth safety. In that setting, ACoFi becomes more conservative when actual safety is inferred to be decreasing, which results in a long-term high probability of preserving safety. The following remarks elaborate on ACoFi's use cases and advantages.

\begin{remark}
    The 0-sublevel set of $V_\theta$ approximates the set of states starting from which and following the policy $\pi^\mathit{safe}_\theta$ leads to reaching $\mathcal{F}$, which in itself is an over-approximation of the BRT of $\mathcal{F}$. While ACoFi does not guarantee preventing
    actions that lead to  $V_\theta(y_{t+1}) < 0$ at some time instances, that does not 
    necessarily imply that following $\pi^\mathit{safe}_\theta$ starting from the time step $t+1$ lead to $\mathcal{F}$, i.e., failure is not inevitable and recovery is possible. The reasons are that $V_\theta$ is only an approximation of the safety value function of $\pi^\mathit{safe}_\theta$. Moreover, the latent state  $y_{t+1}$ might correspond to multiple true states of the system and $V_\theta(y_{t+1})$ is not necessarily the worst value of $\pi^{\mathit{safe}}_\theta$ starting from any such states. Consequently, it is possible that $V_\theta(y_{t+1}) < 0$ and following $\pi^\mathit{safe}_\theta$ from $t+1$ onward prevents reaching $\mathcal{F}$. 
\end{remark}
\begin{remark}
    When a sequence of tasks share the same source of uncertainty for the learned HJ function, ACoFi can carry this adaptation across tasks and keep refining the switching threshold for that specific uncertainty pattern. In that regime, early tasks effectively calibrate the filter for later ones, so the practical guarantee of respecting the safety margin $V_\theta(y)>\epsilon$ approaches the user-chosen threshold over the aggregate deployment horizon rather than only within a single finite-time task. The user-defined threshold $\epsilon>0$ acts as a robustness layer 
    to decrease the possibility of instances at which $V_\theta(y_{t+1})<0$. 
\end{remark}

\section{Experiments}
\label{sec:experiments}
We consider two case studies in this work: First, (1) a vision-based Dubins car setup where an agent with Dubins car dynamics must reach a goal while avoiding two obstacles, and (2) Safety Gymnasium's \texttt{SafetyCarGoal2-v0} environment \cite{ji2023safety}. Both environments are illustrated in Figure~\ref{fig:comparison}. For each case study, we first collect a dataset using a nominal policy. Then, we train a DINO-WM world model using the collected dataset. Next, we derive a HJ value function $V$ from learning the Q-function with (\ref{eq:safety_q_loss}). Finally, we implement the ACoFi algorithm and compare its performance in completing tasks while maintaining safety against the baseline. 
\subsection{Dubins car with Dino-WM}
\label{sec:dubins}
\begin{figure}[h]
    \begin{minipage}[c]{0.24\linewidth} 
        \centering
        \includegraphics[width=\linewidth]{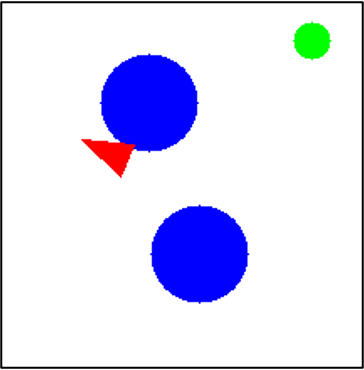}
    \end{minipage}
    \hfill
    \begin{minipage}[c]{0.24\linewidth} 
        \centering
        \includegraphics[width=\linewidth]{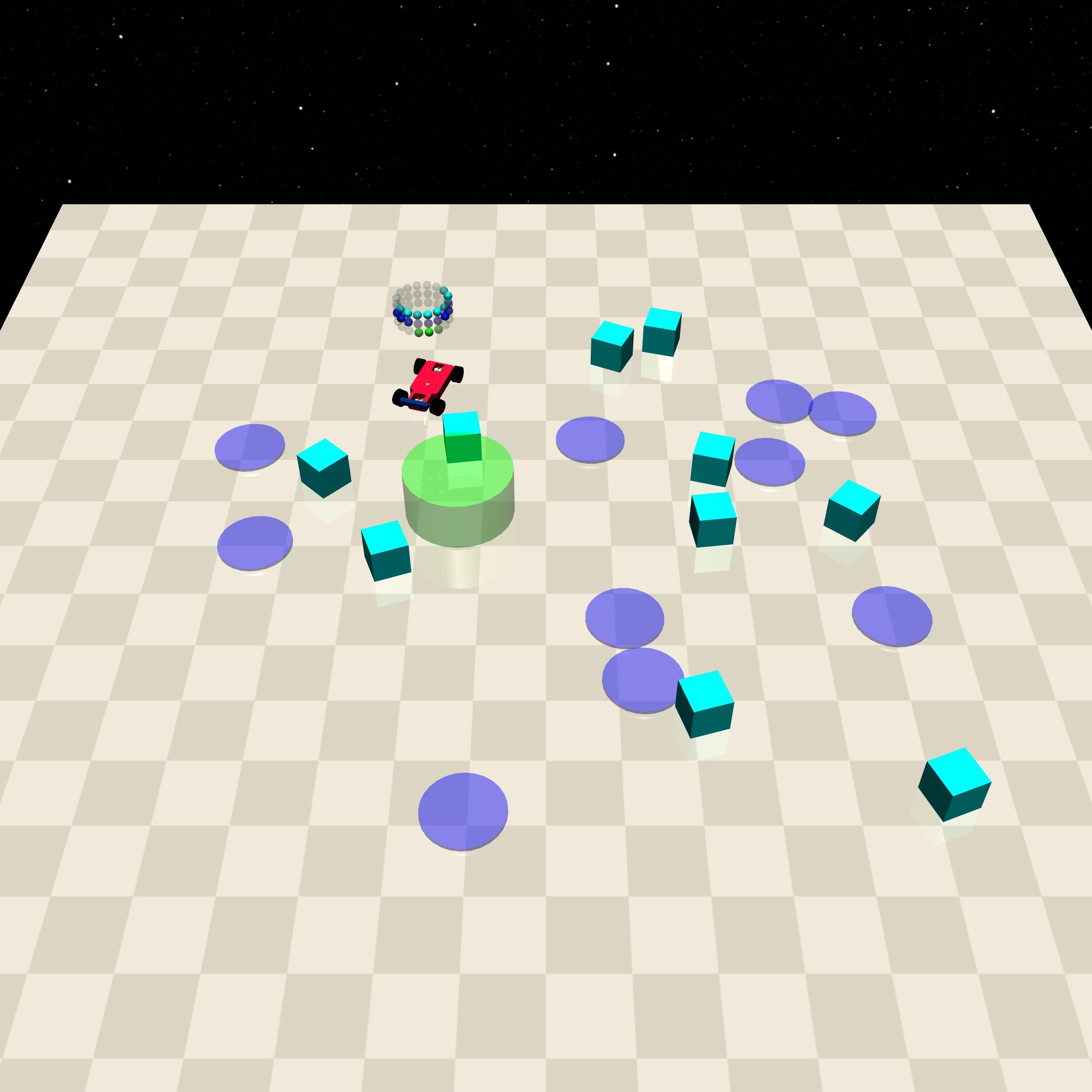}
    \end{minipage}
    \hfill
    \begin{minipage}[c]{0.5\linewidth}
        \caption{\small(Left) The Dubins car is depicted with a \textcolor{red}{red} triangle, with the heading angle in the direction of the narrower corner, the obstacles as \textcolor{blue}{blue} circles and the goal as a \textcolor{green}{green} circle. Violations only happen when the center point of the triangle is inside the obstacle.
        (Right) SafetyGymnasium's Car agent needs to navigate to the goal (\textcolor{green}{green} circle), while avoiding to collide with fixed obstacles (\textcolor{blue}{blue} circles) and with movable obstacle cubes (\textcolor{cyan}{cyan}). 
    }\label{fig:comparison}
    \end{minipage}
    \vspace{-0.2in}
\end{figure}

We first evaluate our approach in a simulated discrete-time 2D Dubins car environment, a controlled benchmark that enables a clear analysis of how our adaptive safety filtering handles uncertainty.
In this experiment, we simulate a car with position $(p_x,p_y)$ and heading $\theta$ inside a bounded space populated with one goal zone and two obstacles. During training, the car moves deterministically at a constant speed $v^{ID}=v$ with the only control being one of three steering actions: $\omega^{ID}_{t}\in\{-\omega,0,\omega\}$, where $\omega=0.05$ rad/step, for setting the angular velocity. The observations are bird-eye view pictures of the environment, as seen in Figure \ref{fig:comparison}. During evaluation, $\pi^\mathit{task}$ is a  PID controller that steers the agent towards the goal, without any consideration for the obstacles. Each run consists of reaching the goal in the top right of the environment five times before the timeout. When the goal is reached or a wall of the environment is hit (not the obstacles), the agent is placed in a random starting position in the lower left.
We discuss the data collection, training the DINO-WM world model, and training the HJ value function in the learned latent space in Appendix \ref{sec:dubins_training}.

\paragraph{Out-of-distribution dynamics}
Uncertainty is simulated by disturbing the dynamics model's parameters at runtime, making the dynamics stochastic. That is:
\begin{gather}
v^{OOD}_{t}\sim v^{ID}+U\left(-1,1\right)\cdot v\qquad\text{and}\qquad\omega^{OOD}_{t}\sim\omega^{ID}_{t}+U\left(-1,1\right)\cdot\omega,\label{eq:ood_dyns}
\end{gather}
where $U\left(-1,1\right)$ is the uniform distribution over the interval $[-1,1]$. 
Since the training was done with constant speed $v^{ID}$ and deterministic action $\omega^{ID}_{t}$, this setup simulates an agent trained on known average values of the dynamics' parameters, and later confronted with large uncertainty on these parameters at runtime due to unmodeled parts of the environment. The HJ value function $V_\theta$ is considered the best estimate of safety the agent has access to before deployment in that environment and ACoFi's role is to adjust decision-making to the observed safety-relevant disturbances to the dynamics.

Using the dynamics in (\ref{eq:ood_dyns}), the four scenarios we consider are {\bf ID:} using the same dynamics as in the Q-function's training, {\bf VarSpeed:} using $v^{OOD}_t$, i.e., a perturbation of the speed in $[-v,v]$ at every step, {\bf VarSteer:} using $\omega^{OOD}_t$, i.e., a perturbation of the control input in $[-\omega,\omega]$, and {\bf VarSpeed\&Steer:} using $v^{OOD}_t$ and $\omega^{OOD}_t$, i.e., both perturbations at the same time.

\paragraph{Baseline}
ACoFi is evaluated against the fixed threshold switching policy $\pi^\mathit{fixed}$. The range of HJ values after training is $[-7.5,+27.5]$, the chosen safety value threshold is $\epsilon=0.1$. We observe that this $\epsilon$ best balanced safety and goal reaching performance when using $\pi^\mathit{fixed}$.
The metrics for comparing adaptive and fixed-threshold switching are the averages over 16 runs of the following measures: the {\em success rate} of reaching the goal $r_\mathit{goal}$, the {\em minimum learned safety value} encountered $\min_t V_\theta(y_t)$, the {\em proportion of steps of violations} $p_\mathit{unsafe}$ of the constraint $V_\theta(y)>\epsilon$, and the {\em proportion of steps using the safe policy} $p_{\pi^\mathit{safe}_\theta}$.
The runs were capped at 1000 steps and are seeded so that the agents using $\pi^\mathit{fixed}$ and ACoFi experience the same OOD disturbances on their speed/steering for fair comparison. The experiments use a target miscoverage rate of $\alpha=0.2$ and a conformal learning rate of $\lambda=0.05$.
\begin{figure}[h]
    \begin{minipage}[c]{0.55\linewidth} 
        \centering
        \includegraphics[width=\linewidth]{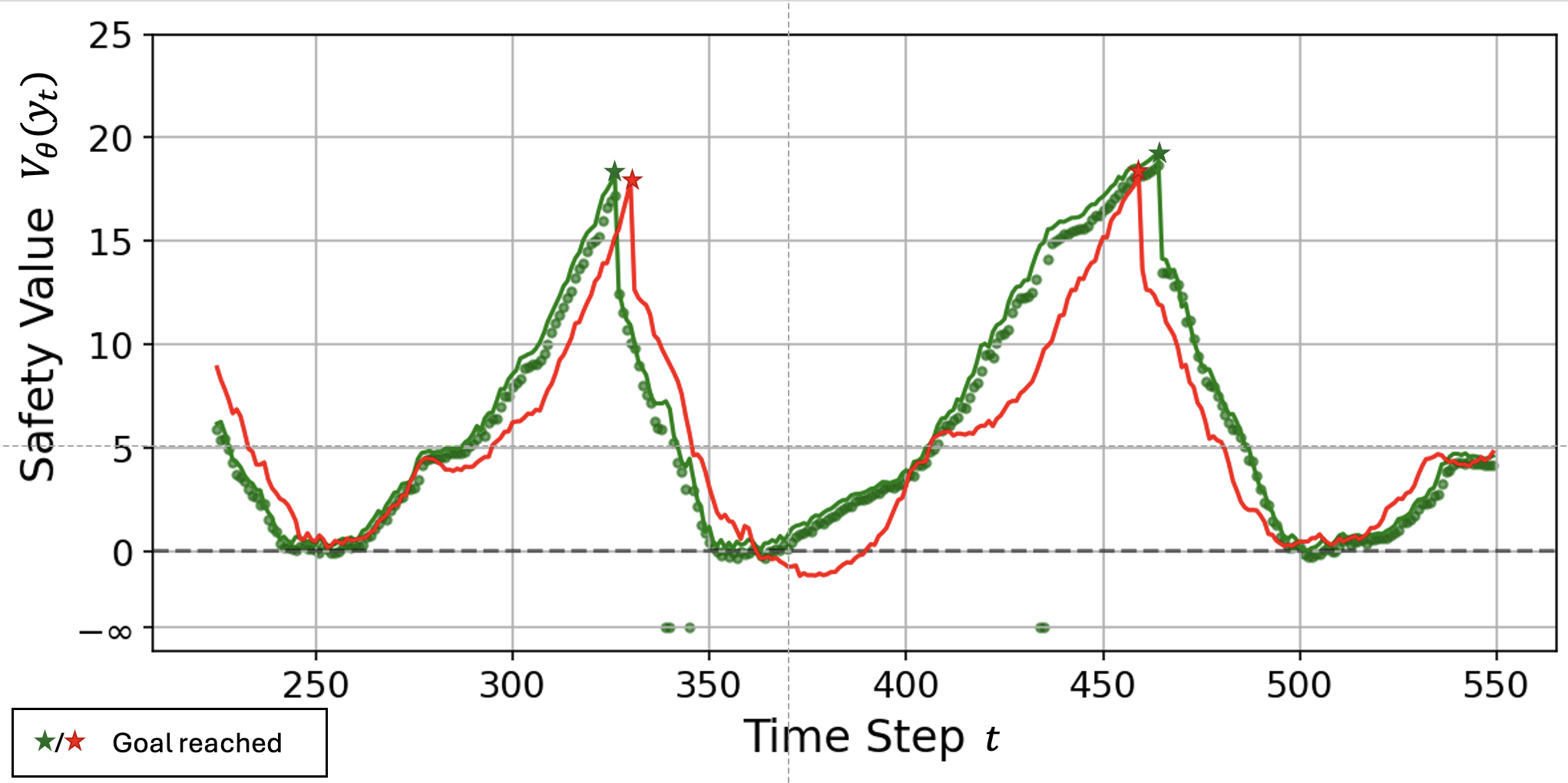}
    \end{minipage}
    \hfill
    \begin{minipage}[c]{0.4\linewidth} 
        \caption{\small Graphs of $V_\theta$ for Dubins car agents using $\pi^\mathit{fixed}$ \textcolor{red}{(red)} and ACoFi \textcolor{darkgreen}{(green)}, under the same {\bf VarSpeed\&Steer} OOD scenario, with safety threshold $\epsilon=0.1$ \textcolor{gray}{(gray)}. The selected time frame shows both agents completing two goal-reaching tasks and being put back in a starting position afterwards. The \textcolor{darkgreen}{circle markers} plot the lower bound $B_{t}$, which is sometimes set to $-\infty$ forcing a switch to $\pi^\mathit{safe}_\theta$ in the case of ACoFi.
    }\label{fig:safety_value_dubins}
    \end{minipage}
    \vspace{-0.15in}
\end{figure}

\subsection{Safety-Gymnasium's CarGoal environment}
ACoFi is evaluated in the CarGoal environment from the Safety-Gymnasium benchmark suite, which provides standardized tests for safe reinforcement learning. In the \texttt{SafetyCarGoal2-v0} environment, the car agent must reach a designated goal region while navigating around two kinds of obstacles: fixed collision regions, and movable cubes (see Figure \ref{fig:comparison}). CarGoal introduces high dimensional state and observation spaces with partial observability, and more complex dynamics compared to the Dubins car task. The goal region, hazard position, and the agent's initial state are randomly sampled at the beginning of every run. The agent is not given access to lidar measurements given by the simulator, which can be used to calculate the distance between the agent and obstacles. We use image observations instead, which makes the task harder. Within the same run, the goal region is resampled once it is reached. The action space is $[-1,1]^2$, representing the force (N) acting on the two independent front wheels. Details on data collection, DINO-WM and HJ value function training are in Appendix \ref{sec:cargoal_training}.
\begin{table}[h]
\centering\footnotesize
    \begin{tabular}{|*{4}{c|}}
    \hline
    \multicolumn{4}{|c|}{\bf ID}\\ \hline
    Policy & $\min_t V_\theta(y_t)$ & $p_\mathit{unsafe}$ & $p_{\pi^\mathit{safe}_\theta}$ \\ \hline
    $\pi^\mathit{task}$ & -1.311 & 15.0/603.4 & - \\
    \hline
    $\pi^\mathit{fixed}$ & 0.040 & 1.2/603.9 & {\bf 10.2/603.9} \\ \hline
    $ACoFi$ & {\bf 0.194} & {\bf 0.0/605.4} & 22.3/605.4 \\
    \hline
    \hline
    \multicolumn{4}{|c|}{\bf VarSpeed}\\ \hline
    $\pi^\mathit{task}$ & -1.346 & 15.2/603.3 & - \\
    \hline
    $\pi^\mathit{fixed}$ & -0.087 & 3.8/608.2 & {\bf 13.9/608.2} \\ \hline
    $ACoFi$ & {\bf 0.158} & {\bf 2.0/605.1} & 22.2/605.1 \\
    \hline
\end{tabular}
\begin{tabular}{|*{4}{c|}}
    \hline
    \multicolumn{4}{|c|}{\bf VarSteer}\\ \hline
    Policy & $\min_t V_\theta(y_t)$ & $p_\mathit{unsafe}$ & $p_{\pi^\mathit{safe}_\theta}$ \\ \hline
    $\pi^\mathit{task}$ & -1.625 & 19.9/605.1 & - \\
    \hline
    $\pi^\mathit{fixed}$ & -0.277 & 4.8/606.3 & {\bf 18.4/606.3} \\ \hline
    $ACoFi$ & {\bf 0.000} & {\bf 1.7/607.7} & 25.7/607.7 \\
    \hline
    \hline
    \multicolumn{4}{|c|}{\bf VarSpeed\&Steer}\\ \hline
    $\pi^\mathit{task}$ & -1.801 & 17.8/607.3 & - \\
    \hline
    $\pi^\mathit{fixed}$ & -0.196 & 3.4/598.8 & {\bf 18.2/598.8} \\ \hline
    $ACoFi$ & {\bf 0.001} & {\bf 2.4/611.6} & 25.7/611.6 \\
    \hline
\end{tabular}
    \caption{\small Results for the Dubins car environment with $\epsilon=0.1$. For both the minimum learned safety value encountered and number of violations, ACoFi ($\alpha=0.2$) is safer than $\pi^\mathit{fixed}$,  and the latter is safer than $\pi^\mathit{task}$.}
    \label{tab:dubins_table}
    \vspace{-0.15in}
\end{table}
\paragraph{Value function learning inaccuracy}
Since the world model and the HJ value function are trained on pre-collected trajectories produced by the Dreamerv3 policy ($\pi^\mathit{task}$), the  distribution of states reached during deployment under a different policy (ACoFi or $\pi^\mathit{fixed}$) is likely to be different than the one reached during training. Thus, the learned HJ value function is likely to be erroneous at some states during evaluation, which is what we observe in our experiments. 
\paragraph{Baselines}
During evaluation, the Dreamerv3 controller serves as the unsafe task policy $\pi^\mathit{task}$. We use it as a baseline against the following policies: $\pi^\mathit{fixed}$ which switches from the task policy to the learned safe policy $\pi_\theta^\mathit{safe}$
whenever the predicted safety value drops below a fixed threshold $\epsilon$, and ACoFi, our proposed method, which replaces that fixed threshold with an adaptive one. We chose $\epsilon$ to be equal to $0.01$ as we observed it best balances safety and task completion when using $\pi^\mathit{fixed}$.
The agent is allowed to navigate in the environment for 1000 steps. The evaluation uses the same metrics as the previous experiment but we replace the success rate $r_\mathit{goal}$ with the {\em number of times the task is achieved} $M_\mathit{goal}$, averaged over all runs like previously. For the target miscoverage $\alpha$, the range of values from $0.1$ to $0.5$ are tested to illustrate its role in our approach. We ran each baseline 25 times with shared seeds, so that they are confronted to the same obstacle and goal locations.

\subsection{Dubins car results}
Figure \ref{fig:safety_value_dubins} illustrates direct comparison between the two switching policies under the same OOD conditions: ACoFi switches to $\pi^\mathit{safe}_\theta$ when the high probability lower bound $B_t$ goes below $\epsilon$, while $\pi^\mathit{fixed}$ switches when $Q_\theta(y_t,\pi^\mathit{task}(y_t))$ does. Over the 16 runs, no agent collided with a wall, and they all reached the goal five times before the timeout, hence $r_\mathit{goal}$ is 100\% for all three baselines. Table \ref{tab:dubins_table} shows the results for the other metrics for all four scenarios. For the two metrics quantifying safety, ACoFi  performs better, with a higher minimum learned safety value and fewer violations than $\pi^\mathit{fixed}$,  which itself improves on $\pi^\mathit{task}$. ACoFi was able to maintain $\min V_\theta(y)>0$ on ID observations and avoid any safety violation over the 16 runs. ACoFi only incurs a minor slow down in goal completion sometimes. In fact, under {\bf VarSpeed}, ACoFi resulted in faster goal reach than $\pi^\mathit{fixed}$ 
on average.
ACoFi results in switches to the safe policy for over twice as many steps as 
$\pi^\mathit{fixed}$ does in the {\bf ID} scenarios. However, for the OOD scenarios, using ACoFi does not incur a stronger reliance on the learned safe policy as much as using $\pi^\mathit{fixed}$ does.
\begin{table}[h]
\begin{minipage}[c]{0.5\linewidth} 
        \centering\small
        \begin{tabular}{|*{5}{c|}}
    \hline
    Policy & $\min_t V_\theta(y_t)$ & $p_\mathit{unsafe}$ & $p_{\pi^\mathit{safe}_\theta}$ & $M_\mathit{goal}$\\ \hline
    $\pi^\mathit{task}$ & -0.294 & 287.2 & - & 5.92 \\
    \hline
    $\pi^\mathit{fixed}$ & -0.107 & 140.1 & 435.0 & 1.16 \\ \hline
    $ACoFi_{\alpha=.5}$ & -0.090 & 54.1 & 438.3 & 0.96 \\
    \hline
    $ACoFi_{\alpha=.4}$ & -0.075 & 44.0 & {\bf 415.8} & {\bf 1.32} \\
    \hline
    $ACoFi_{\alpha=.3}$ & -0.066 & 40.2 & 420.1 & 0.92 \\
    \hline
    $ACoFi_{\alpha=.2}$ & -0.077 & 36.0 & 424.4 & 1.20 \\
    \hline
    $ACoFi_{\alpha=.1}$ & {\bf -0.060} & {\bf 19.9} & 463.4 & 0.80 \\
    \hline
\end{tabular}
    \end{minipage}
    \hfill
    \begin{minipage}[c]{0.475\linewidth} 
        \caption{\small Results for the policies tested in the CarGoal environment. As in the Dubins Car experiment, ACoFi (with $\alpha\leq0.5$) is safer than $\pi^\mathit{fixed}$, and the latter is safer than $\pi^\mathit{task}$, in terms of both the minimum learned safety value encountered and the number of violations of the constraint $V_\theta(y) \geq 0$. Agents using ACoFi see higher safety values and less unsafe states as $\alpha$ decreases, while the average number of goal reaching, which quantifies task completion, is not conclusively decreasing. The total number of steps is omitted since it is 1000 across experiments.}
    \label{tab:cargoal_table}
    \end{minipage}
\end{table}
\subsection{CarGoal results}
ACoFi significantly improves upon the baselines, as Table \ref{tab:cargoal_table} shows. Specifically, it maintained a higher minimum learned safety value and committed up to 7 times fewer safety violations than $\pi^\mathit{fixed}$ on average.
It also did not use the safe policy significantly more than $\pi^\mathit{fixed}$. Finally, it reached fewer goals on average, but only by 30\% at most. Decreasing the target miscoverage rate $\alpha$ steadily improves the minimum learned safety value and reduces the number of unsafe steps, with little effect on the number of steps using the learned safe policy. The tradeoff between safety and task completion seems positive, as $M_\mathit{goal}$ stays close to what $\pi^\mathit{fixed}$ achieved.

\section{Conclusion}
\label{sec:conclusion}
In this work, we aim to enhance the safety of control systems  with high-dimensional observations and relying on learned latent safety filters based on HJ reachability under potential distribution shifts. We propose Adaptive Conformal Filtering, a dynamic extension of the traditional fixed threshold-based switching policy applied within the latent space. Our method was evaluated in two vision-based environments, where it outperformed safety filters which use a fixed switching threshold. Specifically, our adaptive safety filter maintained higher safety values and allowed fewer unsafe actions in out-of-distribution scenarios without significantly increasing reliance on the learned safe policy. Future research should evaluate the effectiveness of this approach for multi-step predictions, and explore an extension to continuous-time control tasks.

\newpage
\section*{Acknowledgments}
This project was partially supported by the NSF CPS award No. 2403758.
\bibliography{l4dc2026-sample}

\newpage
\appendix

\section{Dubins task training}
\label{sec:dubins_training}
 The dataset of trajectories consisting of the RGB images and actions was collected by simulating the agent using a random policy. After collecting this dataset, we train DINO-WM \cite{zhou2025dinowm}. We note that the dataset we used to train DINO-WM does not contain $p_x,p_y,\theta$ as they can be inferred from the image observation. Next, we train the HJ value function in the latent space by optimizing the loss in (\ref{eq:safety_q_loss})  using DDQN. We use $\gamma=0.98$ and define $l(y)$ as the ground truth distance between the agent and the closest obstacle, which we have access to from the simulator. 
\section{Cargoal task training}
\label{sec:cargoal_training}
We collect a dataset of 2000 trajectories using a reward-driven unsafe nominal policy trained with Dreamerv3 \cite{dreamerv3}. We use this dataset to train the DINO-WM. Then, we train the HJ value function by minimizing (\ref{eq:safety_q_loss}) using DDPG. We use $\gamma=0.98$ and assume that $l(y)$ is the ground truth distance function between the agent and the closest obstacle, which we have access to from the LIDAR measurements given by the environment.

\section{Proof of Theorem~\ref{thm:algorithm_guarantees}}
\label{sec:theorem_proof}
Here is the proof of Theorem~\ref{thm:algorithm_guarantees}. 
\pf
    For time step $t\geq 1$, using the fact that $q_{t}\geq0$, we have:
    \begin{align*}
        err_t&=1[s(Q_\theta(y_{t},u_{t}),R_{t})>q_{t}]=1[\max\{Q_\theta(y_{t},u_{t})-R_{t},\ 0\}>q_{t}]=1[Q_\theta(y_{t},u_{t})-R_{t}>q_{t}]\\
        &=1[Q_\theta(y_{t},u_{t})-q_{t}> (1-\gamma)l_{t}+\gamma \min\left\{l_{t},V_\theta(y_{t+1})\right\}],
    \end{align*}
    which implies $err_t\geq1[Q_\theta(y_{t},u_{t})-q_{t}>(1-\gamma)l_{t}+\gamma V_\theta(y_{t+1})]=1[B_{t}>V_\theta(y_{t+1})]$.
    Consequently, we have $1[V_\theta(y_{t+1})\geq B_{t}]\geq1-err_t$, and, by averaging over $1\leq t\leq T$ and using Theorem \ref{thm:conformal_bound}, we get:
    \begin{align*}
        \frac1T\sum_{t=1}^T1[V_\theta(y_{t+1})\geq B_{t}]\geq1-\frac1T\sum_{t=1}^Terr_t=1-\alpha+O\left(\frac1T\right).
    \end{align*}
\qed

\end{document}